\documentclass[12pt]{article}
\pdfoutput=1
\usepackage{amsmath,amssymb}
\usepackage{graphicx}
\usepackage[pdftex]{hyperref}
\numberwithin{equation}{section}
\def\be{\begin{equation}}
\def\ee{\end{equation}}

\def\bea{\begin{eqnarray}}
\def\eea{\end{eqnarray}}

\title{On the Ricci dark energy model}
\author{\large Kyoung Yee Kim, Hyung Won Lee, and Yun Soo
Myung}
\begin{document}
\maketitle

\begin{abstract}
\noindent  We study the Ricci dark energy model (RDE) which was
introduced as an alternative to the holographic dark energy model.
We point out that an  accelerating phase of the RDE is that of a
constant dark energy model. This implies that the RDE may not  be
a new model of explaining the present accelerating universe.


\end{abstract}

\newpage
\section{Introduction}
\noindent Holographic principle~\cite{Bou} regards black holes as
the maximally entropic objects of a given region and implies the
Bekenstein entropy bound of $S_\lambda=(L\lambda)^3 \le S_{BH}=\pi
M_p^2L^2$ with $M^{-2}_p=G$, where $\lambda$ and $L$ are the UV
cutoff and IR cutoff of the given system~\cite{Bek}. On the other
hand, Cohen et al.~\cite{Coh} suggested that the total energy of
the system should not exceed the mass of the same-size black hole,
$E_\lambda=\rho_{\lambda} L^3 \le E_{BH}=LM_p^2$ with the energy
density $\rho_{\lambda}=\lambda^4$. This implies the maximum
entropy bound of  $S_\lambda \le S^{3/4}_{BH}$~\cite{'tH}. Based
on the energy bound, Li has proposed the holographic dark energy
density~\cite{Li}
\begin{equation}
\rho_{\rm L}=\frac{3c^2 m^2_{\rm p}}{L^2}
\end{equation}
with a parameter $c$. Even though we got the holographic energy
density, it is not guaranteed that the holographic energy density
could describe the present accelerating universe.  Hereafter we
choose the reduced Planck mass $m^2_{\rm p}=1/8\pi G=1$ for
simplicity.
 In order for the holographic energy
density to describe the accelerating universe, we have to choose
an appropriate IR cutoff $L$. For this purpose, we may introduce
three length scales of the universe: the apparent horizon (=Hubble
horizon for flat universe), particle horizon, and future event
horizon. The equation of state is defined by
\begin{equation} \label{EOS}
w_{\rm L}=-1-\frac{1}{3 \rho_{\rm L}}\frac{d\rho_{\rm L}}{dx}
\end{equation}
 with $x=\ln a$ the scale factor $a$.

When the cold dark matter (CDM) is present, the Hubble horizon
$L_{\rm HH}=1/H$ does not describe the accelerating universe
because its equation of state $w_{\rm HH}=0$ is the same as the
CDM does show. Here $H=\dot{a}/a$ is the Hubble parameter. The
first Friedmann equation with $\rho_{\rm HH}=3c^2H^2$ leads to \be
\Big(1-c^2\Big)H^2=\frac{\rho_{\rm m}}{3} \ee
 with $\rho_{\rm m}=\rho_{\rm m0}/a^3=\rho_{\rm m0}e^{-3x}$. This provides
$\rho_{\rm HH} \propto 1/a^3=e^{-3x}$, which implies $w_{\rm
m}=0=w_{\rm HH}$~\cite{hsu}.  This means that the holographic dark
energy model with $L_{\rm HH}$ does not describe an accelerating
phase except the interacting case with the CDM~\cite{hor}.

Recently, it was proposed that the energy density which is
proportional to Ricci scalar could derive an accelerating
universe~\cite{gao,rde}. However, its origin is still enigmatic
because it was not proven how an accelerating phase emerges  from
the Ricci dark energy model. Especially, the role of
$\dot{H}$-term is not clearly understood to obtain an accelerating
phase. On the other hand, the $\Lambda$CDM model based on the
cosmological constant is a promising candidate to explain the
present accelerating universe, even though there exist a lot of
dark models.

In this work we show that an  accelerating phase of the RDE is
that of a constant dark energy model with a constant $\omega_{\rm
\Lambda}$. This means that one could not know what kind of dark
energy derives our accelerating universe in the RDE.

\section{ Ricci dark energy model without dark matter}

Let us start with the following  holographic dark energy density
\begin{equation}\label{eq2}
\rho_X=3\alpha \left(2 H^2+\dot{H}\right)
\end{equation}
\noindent where $\alpha$ is constant to be determined.  The first
Friedmann equation is
\begin{equation}\label{eq3}
H^2=\frac{1}{3}\rho_X.
\end{equation}
We can rewrite the first Friedmann equation as
\begin{equation}\label{eq4}
H^2=\alpha \Big(2H^2+\frac{1}{2}\frac{dH^2}{dx}\Big)
\end{equation}
Solving the homogenous equation (\ref{eq4}) for $H^2$ leads to the
conventional Friedmann equation as
\begin{equation} \label{eqq5} H^2=Ce^{-(4-\frac{2}{\alpha})x} \equiv
\frac{\tilde{\rho}_X}{3}.
\end{equation}
\noindent Substituting  $\tilde{\rho}_{\Lambda}$ into the energy
conservation equation,
\begin{equation}
\tilde{p}_{X}=-\tilde{\rho}_{X}-\frac{1}{3}\frac{d\tilde{\rho}_{X}}{dx}
\end{equation}
we obtain the dark energy pressure
\begin{equation}\label{eqq9}
\tilde{p}_{X}=\Big(1-\frac{2}{\alpha}\Big)\,Ce^{-(4-\frac{2}{\alpha})x}.
\end{equation}
\noindent There are two constants $\alpha$  and $C$ to be
determined in the expressions (\ref{eqq5}) and (\ref{eqq9}). Also
the equation of state $\tilde{\omega}_X$ is defined as
\begin{equation}
\tilde{\omega}_X \equiv
\frac{\tilde{p}_{X}}{\tilde{\rho}_{X}}=\frac{1}{3}\Big(1-\frac{2}{\alpha}\Big).
\end{equation}
For $\alpha <1$, it describes the dark energy-dominated universe
in the future.

\section{Constant dark energy model}
We start with the first and second  Friedmann equations
\begin{equation} \label{fsfeq}
H^2=\frac{1}{3}\rho_\Lambda,~~\dot{H}=-\frac{\rho_\Lambda+p_\Lambda}{2}
\end{equation}
where  $\rho_\Lambda$ and $p_\Lambda$ are  the energy density and
pressure in the energy-stress momentum tensor, respectively
\begin{equation}
T^\Lambda_{\mu\nu}=(p_\Lambda+\rho_\Lambda) U_\mu U_\nu -p_\Lambda
g_{\mu\nu}.
\end{equation}
Then the trace of Einstein equation
$G_{\mu\nu}=T^\Lambda_{\mu\nu}$ leads to
\begin{equation} \label{Rieq}
R=-\rho_\Lambda+3p_\Lambda
\end{equation}
with the Ricci scalar $R=-6(2H^2+\dot{H})$~\cite{AP}. The
conservation law of $\nabla^\mu T^\Lambda_{\mu\nu}=0$ implies
\begin{equation} \label{eoscd}
\dot{\rho}_\Lambda +3H(\rho_\Lambda+p_\Lambda)=0.
\end{equation}
Introducing a ``constant" equation of state parameter
$\omega_\Lambda$ like
\begin{equation} \label{eqs}
p_\Lambda=\omega_\Lambda \rho_\Lambda, \end{equation}
 then one solves
equation (\ref{eoscd}) to give
\begin{equation} \label{rhol}
\rho_\Lambda=\rho_{\Lambda 0}
a^{-3(1+\omega_\Lambda)}=\rho_{\Lambda0}e^{-3(1+\omega_\Lambda)x}
\end{equation}
which is a well-known form for a constant $\omega_\Lambda$.

From Eq.(\ref{Rieq}) together with (\ref{eqs}), we have
\begin{equation}
\rho_\Lambda=\frac{R}{-1+3\omega_\Lambda}=\frac{6}{1-3\omega_\Lambda}\Big(2H^2+\dot{H}\Big).
\end{equation}
Now  comparing  $\rho_\Lambda$ with $\rho_X$ in Eq.(\ref{eq2})
implies
\begin{equation}
\alpha=\frac{2}{1-3\omega_{\Lambda}}
\end{equation}
which determines the constant equation of state as function of
$\alpha$ as
\begin{equation}
\omega_{\Lambda}(\alpha)=\frac{1}{3}\Big(1-\frac{2}{\alpha}\Big).
\end{equation}
Plugging this into Eq.(\ref{rhol}), one finds the energy density
and pressure expressed in terms of $\alpha$
\begin{equation}
\rho_\Lambda=\rho_{\Lambda
0}e^{-(4-\frac{2}{\alpha})x},~~p_\Lambda=\omega_\Lambda(\alpha)
\rho_\Lambda.
\end{equation}
Also we find the correspondence for $\rho_{\Lambda 0}=C$
\begin{equation}
p_{\Lambda}=\tilde{p}_X.
\end{equation}
This means that the RDE without  dark matter is nothing but a
constant dark energy model.

\section{Ricci dark energy model with dark matter}
In this section, we include the dark matter so that the first
Friedmann equation takes the form \be
H^2=\frac{1}{3}\Big(\rho_X+\rho_m\Big). \ee The above Friedman
equation becomes
\begin{equation}\label{eq5}
\frac{dH^2}{dx}+\Big(4-\frac{2}{\alpha}\Big)H^2=-\frac{2}{3\alpha}
\rho_m.
\end{equation}
 Solving the inhomogeneous equation (\ref{eq5}) for $H^2$,
we obtain a new  Friedmann equation
\begin{equation}\label{eq6}
H^2=\frac{\rho_{m0}}{3}e^{-3x}+\frac{\alpha}{2-\alpha}\frac{\rho_{m0}}{3}e^{-3x}+Ce^{-(4-\frac{2}{\alpha})x}\equiv
\frac{\rho_m+\tilde{\rho}_X}{3}
\end{equation}
where $C$ is an integration constant.  Importantly,
$\tilde{\rho}_X$ defined by
\begin{equation} \label{trde}
\tilde{\rho}_X=\frac{\alpha}{2-\alpha}\rho_{m0}e^{-3x}+3Ce^{-(4-\frac{2}{\alpha})x}
\end{equation}
plays the role of a new scaled dark energy density.

\noindent Substituting  $\tilde{\rho}_X$ into the energy
conservation equation,
\begin{equation}\label{eq8}
\tilde{p}_X=-\tilde{\rho}_X-\frac{1}{3}\frac{d\tilde{\rho}_X}{dx}
\end{equation}
we obtain the dark energy pressure
\begin{equation}\label{eq9}
\tilde{p}_{X}=\Big(1-\frac{2}{\alpha}\Big)\,Ce^{-(4-\frac{2}{\alpha})x}.
\end{equation}
 Also the ``dynamical"
equation of state $\tilde{\omega}_X$ is obtained  as
\begin{equation}
\tilde{\omega}_X \equiv
\frac{\tilde{p}_{X}}{\tilde{\rho}_{X}}=\frac{1}{3}\Big(1-\frac{2}{\alpha}\Big)
\frac{1}{1+\frac{\alpha\rho_{m0}}{3(2-\alpha)C}e^{(1-\frac{2}{\alpha})x}}.
\end{equation}
For $\alpha <2$ and $x\gg1$, we have  an approximately constant
equation of state
\begin{equation} \tilde{\omega}_X \approx
\frac{1}{3}\Big(1-\frac{2}{\alpha}\Big),
\end{equation}
which describes the dark energy-dominated universe in the future.

\section{Constant dark energy model with dark matter}
In this section, we include the dark matter so that the first
Friedmann equation takes the form \be
H^2=\frac{1}{3}\Big(\rho_{\Lambda}+\rho_m\Big). \ee The above
Friedmann equation leads to
\begin{equation}\label{5eq5}
\frac{dH^2}{dx}+3\Big(1+\omega_\Lambda\Big)H^2=\omega_\Lambda
\rho_m.
\end{equation}
 Solving the inhomogeneous equation (\ref{5eq5}) for $H^2$,
we obtain a new  Friedmann equation
\begin{equation}\label{5eq6}
H^2=\frac{\rho_{m0}}{3}e^{-3x}+\frac{\rho_{\rm \Lambda
0}e^{-3(1+\omega_\Lambda)x}}{3}\equiv
\frac{\rho_m+\rho_\Lambda}{3}.
\end{equation}
 This equation is compared with  Eq.(\ref{eq5}): two homogeneous
equations are the same if one chooses
$\omega_\Lambda=\frac{1}{3}(1-\frac{2}{\alpha})$, while right hand
sides are different as $\omega_{\Lambda}$ and $-2/3\alpha$.  Here
the ``constant" equation of state $\omega_\Lambda$ is obtained  as
\begin{equation}
\omega_\Lambda \equiv \frac{p_{\Lambda}}{\rho_{\Lambda}}.
\end{equation}
For $\alpha <2$ and $x\gg1$, we have   the dark energy-dominated
universe $\rho \simeq \rho_\Lambda$ in the future.

\section{Discussion}
We consider the case without the dark matter. In this case,  we
note that the relation between energy density and Hubble parameter
(its derivative) is found to be

\begin{equation}
\rho_\Lambda=3H^2=-\frac{R}{1-3\omega_\Lambda}=\frac{6}{1-3\omega_\Lambda}\Big(2H^2+\dot{H}\Big)=
\rho_X,
\end{equation}
which  shows clearly that  the Ricci dark energy model is
recovered from a constant dark energy model. The first equality is
the first Friedmann equation for constant dark energy, the second
one relation indicates the trace of Einstein equation (redundant
one), and the last one is the definition of Ricci dark energy
density for $ \omega_\Lambda=\frac{1}{3}(1-\frac{2}{\alpha})$.
This means
 that the Ricci dark energy model is nothing but a constant
dark energy model.   Especially, the role of $\dot{H}$-term is
clearly understood in the Ricci dark energy model. This appears
because the Ricci scalar equation (\ref{Rieq}) (trace of Einstein
equation) incorporates the first ($H^2$) and second ($\dot{H}$)
Friedmann equations in the single, redundant equation. That is, a
mysterious term of $\dot{H}$ is nothing but the term in the second
Friedmann equation (\ref{fsfeq}). Finally, $\alpha$ is replaced by
a constant $\omega_{\Lambda}$.

In the case with dark matter, two are slightly different because
the energy density $\tilde{\rho}_X$ of Eq.(\ref{trde}) is composed
of two parts: one part evolves like dark matter of $e^{-3x}$ and
another part is dark energy term of $e^{-(4-2/\alpha)x}$ with
$\alpha<1$. As a result, we have the ``dynamical" equation of
state  $\tilde{\omega}_X$, while the ``constant" equation of state
$ \omega_{\Lambda}$ is obtained for the constant dark energy
model. Even though there exists a slight difference between energy
densities $\rho_\Lambda$ and $\tilde{\rho}_X$, the Ricci dark
energy model describes constant dark energy model for $x\gg1$.

In conclusion, the origin of Ricci dark energy model is just the
constant dark energy model.

\section*{Acknowledgments}

\end{document}